\documentclass[prl,twocolumn,preprintnumbers,amsmath,amssymb]{revtex4}

\usepackage{epsfig}
\usepackage{psfrag}
\usepackage{upgreek}

\newcommand{\Figname}{Figure }

\begin{document}

\title{Basins of attraction of a nonlinear nanomechanical resonator}

\author{I.~Kozinsky\footnote{The first two authors contributed equally to this
work.}}\altaffiliation{now at Bosch Research and Technology Center, Palo Alto,
CA 94304}
\author{H.W.Ch.~Postma$^*$}\altaffiliation{now at Physics and Astronomy Dept., California State
University at Northridge, Northridge, CA 91330-8268}
\author {O.~Kogan}
\author {A.~Husain}\altaffiliation{now at International Rectifier, El Segundo, CA 90245}
\author {M.L.~Roukes\footnote{Author to whom correspondence should be
addressed; electronic mail: roukes@caltech.edu}}\affiliation{Kavli Nanoscience
Institute and Condensed Matter Physics 114-36, California Institute of
Technology, Pasadena, CA 91125}
\date{\today}

\begin{abstract}
We present an experiment that systematically probes the basins of attraction
of two fixed points of a nonlinear nanomechanical resonator and maps them out
with high resolution. We observe a separatrix which progressively alters shape
for varying drive strength and changes the relative areas of the two basins of
attraction. The observed separatrix is blurred due to ambient fluctuations,
including residual noise in the drive system, which cause uncertainty in the
preparation of an initial state close to the separatrix. We find a good
agreement between the experimentally mapped and theoretically calculated
basins of attraction.
\end{abstract}

\maketitle

In the last few years the dimensions of mechanical devices have been scaled
deep into the submicrometer regime. This decrease in size has resulted not
only in the increased detection sensitivity of extremely small physical
quantities, such as zeptogram-scale mass \cite{Yang:2006} and single electron
spin \cite{Rugar:2004}, but also in an enhancement of the significance of
nonlinear dynamics in such devices \cite{Postma:2005}. The growing
significance of nonlinearities in high-frequency nanomechanical devices has
consequences not only for fundamental studies of nonlinear dynamics, but also
for advances in sensing applications. It was recently shown that precision of
some experimental measurements on nanoscale can be improved by deliberately
operating the system in the nonlinear regime. For example, a nonlinear
resonator can be employed to suppress amplifier noise in an oscillator circuit
\cite{Yurke:1995}, noise-induced switching between two stable states in a
nonlinear beam resonator enables precision measurement of the resonant
frequency \cite{Aldridge:2005}, and the sensitivity of a resonator for mass
detection can be improved when the resonator is driven into a region of
nonlinear oscillations \cite{Buks:2006}. Finally, in a Josephson junction,
which is dynamically similar to a mechanical resonator in nonlinear regime,
the bistable state of the nonlinear system can be used as a bifurcation
amplifier to perform a non-dissipative, low-back-action measurement of the
phase across the junction \cite{Siddiqi:2004}. When nanomechanical devices
reach the quantum-limited regime \cite{LaHaye:2004}, a nanomechanical version
of such an amplifier could be used for a similar sensitive low-back-action
measurement of the state of a quantum mechanical resonator. Nonlinear response
of nanomechanical resonators could also be used to detect transition from
classical to quantum regime \cite{Peano:2004,Katz:2007}. It is therefore
important to understand nonlinear dynamics of these systems well, so that we
can fully realize their potential in expanding our experimental capabilities.

\begin{figure}[ht]
\centerline{\epsfig{file=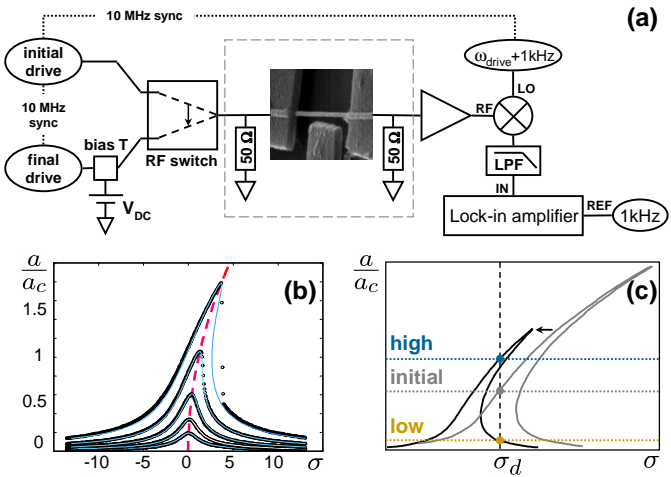, width=9cm, clip=}}
\caption{ \label{fig-setup} {\bf (a)} Experimental layout. The initial
drive prepares an initial state of the platinum nanowire resonator (shown in
the SEM photo), a 5 ns RF switch is then flipped to connect to the final
drive, and the state of the device is measured by a lock-in amplifier after
mixing down to a low (1 kHz) frequency and filtering the residual RF signal.\\
{\bf (b)} Vibration amplitude versus frequency, for various driving powers at
the sample (-90, -85, -80, -75, -70 dBm, or normalized to the critical drive
$V_c$=28.4 \mbox{$\mu$V}, $V/V_c$=0.249, 0.443, 0.788, 1.401, 2.492) showing
the onset of nonlinearity in the platinum nanowire resonator. We plot the
response normalized to the critical amplitude, $a_c$, versus normalized
detuning frequency $\sigma \equiv 2 Q (f/f_0 - 1)$. The backbone curve (dashed
line) connects the maxima of the resonance curves and follows $(a_p/a_c)^2 =
\sqrt{3}\sigma/2$.\\ {\bf (c)} The state of the nanowire resonator is first
prepared in an initial state on the initial response curve (grey) by choosing
an appropriate drive strength for the fixed detuning frequency,
$\sigma_d=4.26$. After the RF switch is flipped to connect the final drive,
the response curve changes to the one shown in black and the initial state
evolves to either the high-amplitude state or low-amplitude state. The small
bias voltage, applied to the nanowire together with the final drive, is chosen
so that the hysteresis loop of the initial curve is at higher frequency than
the operating frequency.}
\end{figure}

Although some work has been done with parametric systems \cite{Cleland:2005,
Karabalin:2007}, the majority of nonlinear nanoscale systems that have been
studied are directly driven \cite{Aldridge:2005,Husain:2003,Erbe:2000} and the
dominant nonlinearity in the restoring force is cubic, also known as Duffing
nonlinearity. When a system is driven strongly, the Duffing nonlinearity
causes the resonance response curve to become asymmetric. The resonance is
pulled either to the right for positive, also known as hardening, nonlinearity
(e.g. geometric nonlinearity \cite{Landau:1986}, \Figname \ref{fig-setup}(b))
or to the left for negative, or softening, nonlinearity (e.g. nonlinearities
of material \cite{Bolotin:1964}, capacitive, or inertial \cite{Atluri:1973}
origins). When the resonance is pulled far enough to one side, hysteretic
behavior is observed as two stable states appear in the system
\cite{Landau:1981}. The stable states, known as "attractors" or "fixed
points", correspond to the points in state space to which the system converges
with time. For each attractor, a set of initial states that dynamically
evolves to that attractor forms its basin of attraction, which is separated
from the rest of the state space by the separatrix curve. The dependence of
fixed points and basins of attraction on driving frequency and amplitude has
been exploited for precision measurement applications mentioned earlier.

There have been very few experimental studies of basins of attraction because
following the evolution of initial conditions in low-frequency macroscopic
systems is usually very time-consuming and system parameters tend to drift
over the course of many data-taking runs. Previous mappings of basins of
attraction \cite{Cusumano:1995, Virgin:1998} used the method of stochastic
interrogation, where the system is stochastically perturbed and initial states
are sampled at random without fully covering the basins.

We demonstrate the ability to systematically prepare a nonlinear, Duffing-type
nanomechanical resonator in the required set of initial states and map the
basins of attraction of its two fixed points with high resolution. Our
experiment is enabled by the fact that the relevant time scale per data point,
$\sim Q/f_0$, is very short for very-high-frequency nanometer-scale devices.
This allows us to take many data points in a relatively short time with
minimal drift in the parameters of the system. We also observe that the
separatrix changes shape for varying drive strength, so that one of the basins
becomes progressively smaller and eventually disappears. The mapped basins of
attraction show good agreement with theory. However, the observed separatrix
is blurred due to ambient fluctuations, including residual noise in the drive
system, which cause uncertainty in the preparation of an initial state.

The device used for mapping the basins of attraction, a doubly-clamped
platinum nanowire, is shown in the scanning electron microscope (SEM)
photograph in \Figname \ref{fig-setup}(a). The nanowire, with a length $L$ of
$2.25\mbox{ $\upmu$m}$ and a diameter of $35\mbox{ nm}$, is grown by
electrodeposition of platinum into a nanoporous membrane \cite{Martin:1994}.
Gold contact pads on both ends and a gate are fabricated using electron beam
lithography, and about 150 nm of the substrate is subsequently etched away in
hydrofluoric acid to suspend the device \cite{Husain:2003}. We actuate and
detect the vibration of the nanowire magnetomotively \cite{Cleland:1999a} in a
magnetic field, $B=8$ T, in a cooled (to about $20$ K) probe in vacuum . The
magnetic field is applied perpendicular to the device so that the vibration is
in the plane of the gate electrode. At low driving powers the resonance curve
is linear, and we extract a resonant frequency of $45.35\mbox{ MHz}$ and
mechanical quality factor of $6045$. The resonant frequency is higher than the
expected $17.71\mbox{ MHz}$ for this device geometry, most likely due to
differential thermal contraction between the silicon wafer and the gold
contacts that results in residual tension. The ratio of electromechanical
impedance, $R_{em}$, to electrical impedance, $R_e$, is $0.222$, which
indicates the presence of significant eddy current damping \cite{Schwab:2002}.
The quality factor corrected for the eddy current damping is $Q_0 =
Q/(1-R_{em}/R_e) = 7770$.

The resonant response of the nanowire to different drives is shown in \Figname
\ref{fig-setup}(b). With increasing drive power, the resonance is pulled to
higher frequencies at large amplitudes, ultimately forming a hysteretic
region. The nonlinearity of the device is fully characterized by the critical
amplitude $a_c$, the point where the resonance curve develops infinite slope,
$\frac{\mathrm{d}a}{\mathrm{d}f}\mid_{a=a_c} =-\infty$. The theoretical curves
(thin solid lines) in \Figname \ref{fig-setup}(b) are generated using the
critical amplitude, $a_c$, as the only fitting parameter. We determine $a_c$
by fitting the backbone curve that connects the peaks of resonant curves for
different drives to the theoretical expression $\left(a_p/a_c\right)^2 =
\sqrt{3} \sigma/2$ \cite{Nayfeh:1979}, where $\sigma \equiv 2Q(f/f_0-1)$ is
the detuning frequency scaled by the width of the resonance and $a_p$ is the
peak amplitude. The critical amplitude value $a_c = 2.68\mbox{ nm}$, extracted
from the experimental data in this manner, is in reasonable agreement with the
value calculated for our nanowire geometry \cite{Postma:2005} when the
round-trip loss in the experimental circuit is taken into account.

The dominant source of nonlinearity in doubly-clamped NEMS resonators is the
additional tension in the beam that appears when vibrations are sufficiently
large. This extra tension gives rise to a cubic nonlinearity in the spring
constant term in the equation of motion \cite{Postma:2005}, giving it a
Duffing oscillator form:
\begin{equation}\label{EOM}
\ddot{x} + \frac{\omega_0}{Q} \dot{x} + \omega_{0}^{2}(x+\alpha x^3) =
F\cos(\Omega t).
\end{equation}
Here, $x(t)$ is the displacement of the beam, $\omega_0=2\pi f_0$ is the
resonance frequency, $\alpha=2\sqrt{3}/(9a_c^2Q)$ is the nonlinearity
parameter \cite{Nayfeh:1979}, $\Omega=2\pi f=(\sigma/(2Q)+1)\omega_0$ is the
driving frequency, and $F$ is the force per unit mass acting on the resonator
of mass $m$. The driving force in the magnetomotive transduction scheme is the
Lorentz force that acts on the nanowire when a current $I_d(t)$ is passed
through it in a magnetic field, $F=LBI_d(t)/m$. When the driving voltage
amplitude is $V_d$, the driving current is given by $I_d \approx V_d/R_e$.

% Experimental setup
As illustrated in \Figname \ref{fig-setup}(a), we prepare the system by
exciting it with an initial drive amplitude, $V_i$, and then switching to a
final drive amplitude, $V_f$. Two RF sources (HP 8648B for initial and SR DS345
with a frequency doubler for final drive) are tuned to the same fixed frequency
off resonance, $\sigma_d=4.26$. Their internal clocks and that of the local
oscillator (LO) are synchronized with their 10 MHz clock reference. The phase
of the final drive lags behind the initial drive phase by the phase difference
$\phi$. By changing $\phi$ while holding the initial drive amplitude, $V_i$,
constant, we can prepare the resonator in the initial states corresponding to a
circle in state space. By also stepping the initial drive values, $V_i$, we can
cover a disk of initial states in state space. We switch rapidly from the
initial to final drive using a 5 ns ($< 1/f_0\sim$22 ns) RF switch
(Mini-Circuits ZASWA-2-50DR). After the switching occurs, we measure the final
amplitude of the oscillator and mark it as a low or a high final amplitude. In
order to access a continuum of initial states, we apply a small DC bias voltage
of $V_{dc}\approx 10$ mV to the wire in the final state. The capacitive
interaction with the gate lowers the resonant frequency of the final state
\cite{Kozinsky:2006} so that the hysteretic frequency response of the final
state occurs at the same frequency as the single-valued resonant response of
the initial state (\Figname \ref{fig-setup}(c)). Without this technique, the
resonator cannot be prepared in the initial states corresponding to the
unstable branch of the initial drive resonance and an annulus of these states
in state space would not be accessible in the experiment.

\begin{figure}[ht]
\centerline{\epsfig{file=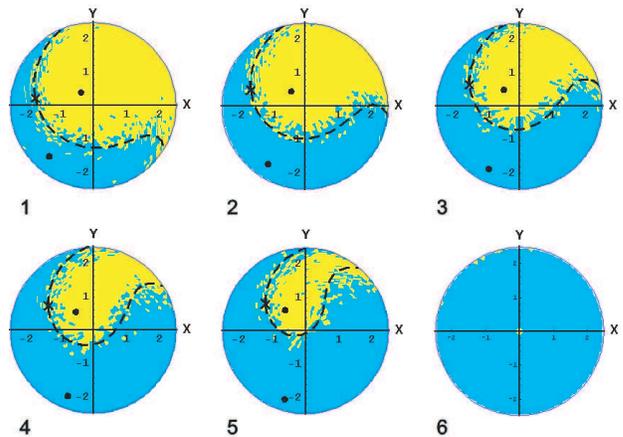, width=9cm}}
\caption{\label{fig-basins} Basins of attraction of a nanowire resonator
at a fixed frequency, $\sigma_d=4.26$, for increasing final drive values,
$V_f/V_c$=$(1)1.867,(2) 2.049, (3) 2.237, (4) 2.434, (5) 2.640, (6) 2.741$.
Blue and yellow colors indicate the final high- and low-amplitude states,
respectively. The data consists of 60 concentric circles with 60 points each,
corresponding to a displacement range of 0 to 2.5$a_c$. This data is converted
into a continuous plot using a nearest neighbor search algorithm to fill out
the space between data points. Theoretical fixed points, saddle point, and
separatrix curve are indicated by black points, black cross, and the dashed
black curve respectively.}
\end{figure}

To map the basins of attraction in this manner, the initial states were driven
with $-90$ to $-50$ dBm in 60 concentric circles with 60 phase points per
circle, corresponding to a displacement range of 0 to 2.5$a_c$, where $a_c$ is
the critical amplitude extracted from experimental data as shown in \Figname
\ref{fig-setup}(b). Each of the initial states was marked according to the
attractor that it evolved to after the switch was flipped: blue for the
high-amplitude state and yellow for the low-amplitude state. This data was
re-rastered using a nearest neighbor search algorithm to create the continuous
color plots shown in \Figname \ref{fig-basins}.

For very low final drives, there is only one state the resonator can occupy
(refer to \Figname \ref{fig-setup}(b)). As the final drive starts exciting a
nonlinear response, a second basin representing the high-amplitude stable
state appears, but subtends a smaller fraction of the state space (\Figname
\ref{fig-basins}(1)). When the final drive amplitude is increased, the
high-amplitude basin grows, and the low-amplitude basin gets progressively
smaller and eventually disappears (\Figname \ref{fig-basins}(6)). The
disappearance of the low-amplitude state can also be observed in \Figname
\ref{fig-setup}(b): a large final drive results in a wider hysteretic region
that moves to the right, where the low amplitude branch starts at
$\sigma>\sigma_d$, so only the high-amplitude state is available to the system
operating at $\sigma_d$.

% Theory
In order to calculate the location of the fixed points and the separatrix in
state space, we obtain the time-dependent solution $x(t)$ that describes the
response of the nonlinear system to a disturbance. In our devices, we can
separate the dynamics described by the equation of motion (\ref{EOM}) into two
parts: the fast dynamics on a time scale of $1/\omega_0$, corresponding to the
fast oscillations of the undamped harmonic version of the system; and the slow
dynamics on a much longer time scale $Q/\omega_0$, associated with a slight
detuning $\sigma$ of the driving frequency from resonance as well as damping
and nonlinearity (method of multiple scales \cite{Nayfeh:1979}). Then the
solution to the equation of motion (\ref{EOM}) can be written as $x_0(t,T) =
A(T)e^{i\omega_0 t} + \bar{A}(T)e^{-i\omega_0 t}$, where time variable $t$
characterizes the fast dynamics and $T$, the slow dynamics. The slowly varying
amplitude $A(T)=(X(T)+iY(T))\exp(i\omega_0\sigma/2T)$ obeys the envelope
equations:
\begin{eqnarray}\label{XYtheory}
\frac{\mbox{d}X}{\mbox{d}T} &=& -\frac{\omega_0 X}{2}+\frac{\sigma\omega_0}{2} Y-\frac{3\alpha Q\omega_0}{2}(X^2 + Y^2)Y\\
 \frac{\mbox{d}Y}{\mbox{d}T} &=& -\frac{\omega_0
Y}{2}-\frac{\sigma\omega_0}{2} X+\frac{3\alpha Q\omega_0}{2}(X^2 +
Y^2)X-\frac{FQ}{4\omega_0}.\notag
\end{eqnarray}
We have assumed here that $1/Q\ll1$, and the slowly-varying amplitude
approximation implies that $\ddot{A}(t)$ terms are negligible compared to
$\dot{A}$ terms.

The slowly-varying amplitude equations describe the nonlinear dynamics of the
system and allow us to determine the location of two attracting fixed points
and one metastable saddle point in state space for different values of
parameters $F$ and $\sigma$, which can be extracted from the experimental
data. A set of points in state space that evolve into the saddle point defines
the separatrix. To calculate the separatrix, we evolve the initial conditions,
lying close to this fixed point and along the negative-eigenvalue eigenvector
(which is obtained by linearizing the above equations around the saddle fixed
point), backwards in time according to the full equations (\ref{XYtheory}).
The curves generated by this procedure constitute the separatrix. To compare
these theoretical calculations to the experimental data in \Figname
\ref{fig-basins}, we scale computed amplitudes by the value of the critical
amplitude, $a_c$, calculated from equations (\ref{XYtheory}).
% end theory

The black points in \Figname \ref{fig-basins} are the theoretical fixed
points, the black cross is the saddle point, and the dashed black curve
corresponds to the theoretical calculation of the separatrix for the
experimental parameters used: $\sigma_d=4.26$, $(a/a_c)_{max}=2.5$, and the
first five final drive values scaled by critical drive, $V_f/V_c$, listed in
the caption. We find good agreement between the experimental data and our
theoretical calculations for the basins of attraction.

% V.2
The separatrix observed in the experiment is blurred due to environmental noise affecting the system. The following analysis implies that most of this noise results from the voltage noise in the drive circuit. These fluctuations perturb the initial state of the resonator and so cause uncertainty in the preparation of the initial state. For the states near the separatrix, switching the system to the bistable regime and forcing the initial state to project on either of the two stable states amplifies this jitter and results in the noisy separatrix in \Figname \ref{fig-basins}. In principle, the full solution for the probability distribution of the initial state of the Duffing resonator with noise would be obtained by solving the Fokker-Plank equation. Since the system is operated away from the bifurcation, the time scale on which noise affects the system is much slower than the ringdown time of the resonator and the dynamics can be analyzed in the small-noise approximation. We can then consider two scenarios for how external noise can affect the system, either additively or as parameter noise.

If noise in equations (\ref{XYtheory}) for the initial state is additive, the evolution of the initial condition follows the deterministic path perturbed by small jitter until the system reaches the neighborhood of an attracting fixed point, corresponding to the initial state. The action of this noise is to perturb the system around the fixed point within a characteristic radius $\delta a$. If the additive noise in equation (\ref{EOM}) is white with power spectral density $\gamma$, then the effective temperature of the noise in the system described by equations (\ref{XYtheory}) is $\frac{\gamma\alpha Q^2}{8m^2\omega_0^3}$. From the equipartition theorem, $\delta a$ is given by
\begin{equation}\label{std-a2}
\frac{(\delta a)^2}{a_c^2}=\left(\frac{3}{\sqrt{2}}\right)^{3/2}\frac{\gamma\alpha Q^2}{4m^2\omega_0^3}.
\end{equation}
We can estimate the uncertainty in the preparation of the initial amplitude from the observed blurring of the separatrix, $\delta a/a_c$, in \Figname \ref{fig-basins} to be about 10\%. The force noise that would result in such blurring is 5~pN/$\sqrt{\mbox{Hz}}$. For the Lorentz force due to the magnetomotive drive, this force noise translates to a voltage noise of 18~nV/$\sqrt{\mbox{Hz}}$.

Alternatively, the noise can be present in the parameters of the system. The noise in the magnitude of the drive, $F$, results in the fluctuations of the position of the initial fixed point. Near the origin, where $a/a_c$ is small, the variation in the drive voltage $\delta V/V_c$ that would cause this variation in amplitude is calculated from equations (\ref{XYtheory}) to be $\delta V/V_c=(\sqrt{3}/2)a/a_c$. For the estimated blurring of the separatrix, $\delta a/a_c$, of about 10\%, the variation in the drive amplitude is $\delta V/V_c$ = 8.7\%. (Farther away from the origin, up to $a/a_c = 2.5$, fluctuations in the drive voltage have less effect on fluctuations in the amplitude due to nonlinear suppression.) Therefore, the observed blurring translates to fluctuations of $\delta V$ $\approx$ 2.5~$\mu$V for the measured
critical drive, $V_c$ = 28~$\mu$V. The relevant noise bandwidth for this
resonator is $\pi f_0/Q$ = 24~kHz. The noise spectrum that would account for the
10\% fluctuation in amplitude is then 16~nV/$\sqrt{\mbox{Hz}}$.
This result is consistent with the above estimate of fluctuations due to additive noise. The residual voltage noise from the initial-drive function generator and the rest of the drive circuit of about 5 $\mu$V/$\sqrt{\mbox{Hz}}$ is attenuated by 51~dBm by the RF switch and
additional attenuators (not shown), and results in a voltage noise of 14~nV/$\sqrt{\mbox{Hz}}$ at the sample. The drive-circuit noise therefore accounts for most of the observed fluctuations near the separatrix.

% noise transitions
The presence of noise can also cause transitions from one fixed point to
another \cite{Kurkijarvi:1972,Dykman:1979}. We indeed observe the same
noise-induced switching between two stable states as in References
\cite{Aldridge:2005} and \cite{Stambaugh:2006prb}, where the noise from the
drive source has an effect of shrinking the size of the hysteresis loop and
inducing transitions of a resonator from one state to the other near the
bifurcation points. By adding noise to the resonator drive and recording the
statistics of the time it takes for the system to switch when it is near the
bifurcation point, we find that the transition rate varies as
$\exp(-E_a/\nu)$, where $\nu=k_BT_{eff}$ is the noise power and $E_a$ is the
height of the energy barrier that the system needs to overcome for the
transition to happen. The energy barrier depends on the distance to the
bifurcation point $V_b$: $E_a\sim(V-V_b)^\delta$. We have measured the
critical exponent $\delta$ to be $1.8\pm 0.3$, which is close to the
theoretically predicted value of $3/2$ in the region we operate in
\cite{Kurkijarvi:1972,Dykman:1979,Kogan:2007}. Measurement of transitions
induced by noise in the bistable regime of a nanoscale resonator could thus
enable a very sensitive experimental technique to detect mechanical
fluctuations.

The experimental mapping of basins of attraction of a nanowire mechanical
resonator presented here fills a large gap in our understanding of nonlinear
dynamics of nanoscale systems. Since the nonlinear regime is readily
accessible in nanoscale devices, the details of dynamical behavior are now
increasingly important for proper engineering and analysis of these systems.
The knowledge of basin dynamics and noise-induced transitions should also
prove useful for precision measurement applications, such as nanomechanical
bifurcation amplifiers \cite{Karabalin:2007}, detection of transition to the
quantum regime \cite{Peano:2004,Katz:2007}, or a sensitive monitoring of
intrinsic device noise processes.

We gratefully acknowledge partial support from the DARPA MTO/MGA (via DOI
NBCH1050001). O.~Kogan acknowledges support from the NSF grant DMR-0314069. We
thank M.~Barbic for building the nanowire fabrication setup and I.~Bargatin
and M.C.~Cross for useful comments on the manuscript.

%\bibliographystyle{apsrev}
%\bibliography{references}

%\widetilde{}\newpage
%\textbf{Bibliography}

\end{document}